\documentclass[a4paper]{article}

\usepackage{amssymb}
\usepackage{latexsym}

\usepackage{url}
\usepackage{xcolor}
\usepackage{graphicx}

 \usepackage{listings}
\definecolor{mygray}{rgb}{0.5,0.5,0.5}
\newcommand{\eg}{\emph{e.g.}, }       
\newcommand{\ie}{\emph{i.e.}, }      
\newcommand{\etal}{\emph{et al.}}         
\newcommand\etc{\emph{etc.}}

\newtheorem{definition}{Definition}

\usepackage[ruled]{algorithm2e}
\renewcommand{\algorithmcfname}{ALGORITHM}
\SetAlFnt{\small}
\SetAlCapFnt{\small}
\SetAlCapNameFnt{\small}
\SetAlCapHSkip{0pt}
\IncMargin{-\parindent}

\usepackage[style=numeric,backend=bibtex]{biblatex} 
\bibliography{thesis_bib}

\title{Hierarchical Hyperlink Prediction for the WWW}

\author{Garcia-Gasulla, D. \& Ayguad\'{e}, E. \& Labarta, J.\\
Barcelona Supercomputing Center (BSC)\\dario.garcia@bsc.es\vspace{8pt}\\
\& Cort\'{e}s, U.\\
Universitat Polit\`{e}cnica de Catalunya - BarcelonaTECH\vspace{8pt}
\& Suzumura, T.\\
IBM T.J. Watson Research Center\\Barcelona Supercomputing Center (BSC)\vspace{8pt}
}

\begin{document}

\maketitle

\begin{abstract}
The hyperlink prediction task, that of proposing new links between webpages, can be used to improve search engines, expand the visibility of web pages, and increase the connectivity and navigability of the web. Hyperlink prediction is typically performed on webgraphs composed by thousands or millions of vertices, where on average each webpage contains less than fifty links. Algorithms processing graphs so large and sparse require to be both scalable and precise, a challenging combination. Similarity-based algorithms are among the most scalable solutions within the link prediction field, due to their parallel nature and computational simplicity. These algorithms independently explore the nearby topological features of every missing link from the graph in order to determine its likelihood. Unfortunately, the precision of similarity-based algorithms is limited, which has prevented their broad application so far. In this work we explore the performance of similarity-based algorithms for the particular problem of hyperlink prediction on large webgraphs, and propose a novel method which assumes the existence of hierarchical properties. We evaluate this new approach on several webgraphs and compare its performance with that of the current best similarity-based algorithms. Its remarkable performance leads us to argue on the applicability of the proposal, identifying several use cases of hyperlink prediction. We also describes the approach we took for the computation of large-scale graphs from the perspective of high-performance computing, providing details on the implementation and parallelization of code.
\textit{Web Mining, Link Prediction, Large Scale Graph Mining}
\end{abstract}

\section{Introduction}\label{intro}

Link prediction is a link mining task focused on discovering new edges within a graph. Typically, we first observe the relationships (or \emph{links}) between some pairs of entities in a network (or \emph{graph}) and then we try to predict unobserved links \cite{Miller-NIPS2009}. Link prediction methods are relevant for the field of Artificial Intelligence (AI) as for any Knowledge Base (KB) represented as a graph these methods may produce new knowledge in the form of relations. In most cases link prediction methods use one of two different types of graph data for learning \cite{LPCN,LPSP}: the graph structure (\ie its topology) and/or the properties of the graph components (\eg node attributes). Combinations of both have also been proposed \cite{DLP}.

Solutions based on the properties of components must rely on a previous study of those properties (\eg which attributes are present, which is their type, their relevance, \etc). These solutions are therefore only applicable to domains where those properties hold, and can hardly be generalized. On the other hand, solutions based on structural information of the graph are easily applicable to most domains as this information is found in all graphs and their interpretation is broadly shared (\ie all graphs have a topology and its interpretation is the connectivity of elements). In this article we will focus on link prediction methods based on structural data.

The problem of link prediction is commonly focused on undirected, unweighted graphs. Whereas plain edges have some universal semantics (\ie the existence of a relationship between two entities), additional properties such as directionality and weights may have varying interpretations (\eg what is the difference between the origin and the destination of an edge? what is the impact of weights?). Link prediction methods taking into account one or more of these additional properties must make certain assumptions regarding their semantics, assumptions which will limit the domains to which they can be applied. However, if the semantics of these properties are shared by a large set of domains, the development of specific link prediction methods with these assumptions may be beneficial; these methods may increase the quality of the results while still being applicable to a relevant set of domains.

\subsection{Hyperlink Prediction}

The World Wide Web naturally represents a \emph{webgraph}: a large, directed graph composed by web pages connected through hyperlinks. In contrast with traditional instance-attribute based data sets, webgraphs define highly distributed and inter-connected structures \cite{WEBHIER,HIERNET}, often referred to as networks. Networks became popular data structures with the explosion of the Internet and motivated a change on how data was being processed by fields like Data Mining (DM) and Machine Learning (ML), as these fields had to shift their focus towards finding \emph{inter-entity} patterns (\ie entity-entity relations). Graph-based data mining \cite{GDM}, Statistical Relational Learning \cite{SRL}, Link Mining \cite{LPCN}, Network or Link Analysis \cite{GUI}, Network Science \cite{NEWPERS} or Structural Mining \cite{STRMIN} are all names used to identify these knowledge discovery tools which share the same precept: exploiting structural properties of high-dimensional, inter-connected data sets to learn about the relational patterns of its entities. For the sake of simplicity, we generally refer to this field of science with the term \emph{graph mining}. 

A defining contribution to the graph mining field was the identification of certain problems which had not been successfully tackled by traditional DM and ML techniques \cite{LINK}. Tasks like community detection \cite{STOCH}, frequent subgraph discovery \cite{APRIORI} and link-based object ranking \cite{PR}. These were to become the main target of graph mining algorithms. One of those tasks, the link prediction (LP) problem of finding new relations in a high-dimensional data set (\eg adding new edges to a given graph), is one with a particularly wide range of applications. Through LP one can automatically produce new knowledge (edges) within a domain (a graph) using its own language (the topology of vertices and edges), avoiding any interpretation step. This infrequent and enabling feature should have made of LP a technique widely applied to domains that could be represented as a graph. Unfortunately scalable LP methods are often either imprecise or non-generalizable, problems that have constrained their popularization so far.

A prototypical domain of application of LP is the WWW webgraph, a large and sparse graph growing continuously in a distributed manner. Any algorithm to be applied to large graphs such as webgraphs needs to be scalable. In fact, their size often motivates the use of High Performance Computing (HPC) tools and infrastructure, even when using scalable methods. The challenges arising from the computation of large networks has been noticed by the HPC community, as demonstrates the popularization of the Graph500 benchmark to evaluate on how fast can supercomputers process large-scale graphs \cite{TOYO1}, or the development of graph-specific parallel programming models \cite{PREGEL}. The collaboration between graph mining and HPC seems therefore to be inevitable in the near future, as both fields may benefit from it: graph mining by obtaining the means for its goals and HPC by obtaining a purpose for its capabilities. In this context, the work here presented provides the following contributions:

\begin{itemize}
 \item WWW: Discuss the importance of hierarchical properties for the topological organization of the WWW. Study the performance of four different LP scores for the specific problem of hyperlink prediction, including one that exploits hierarchical properties.

 \item Graph mining: Propose an improved version of \emph{INFerence}, a score for directed link prediction based on hierarchies, to process informal graphs not explicitly hierarchical. Compare the results with the current best similarity-based algorithms. Study the distinct types of local LP algorithms, and proposed a combined solution to significantly improve performance.

 \item HPC: Present a feasible and applicable use case of large-scale graph mining. Discuss the scalability and parallelism of similarity-based LP algorithms and how these can be extended to distributed memory setting for the computation of large-scale graphs.
\end{itemize}

The rest of the paper is structured as follows: the first two sections survey related work. \S\ref{sec:LP_back} summarizes the current state-of-the-art on LP, and contains a brief survey on previous attempts to perform LP on WWW data. \S\ref{sec:web_hier} reviews the role of hierarchies in the WWW according to previous work. Once the related work has been presented, in \S\ref{sec:tunn} we introduce and formalize the Hybrid\-INF score, a novel LP algorithm for the task of hyperlink prediction. We describe our experiments in \S\ref{sec:experiments}, including the webgraphs used, the evaluation methodologies and the analysis of results. From those results we derive various possible applications in \S\ref{sec:app}. We discuss computational details in \S\ref{sec:large}, describing our implementation and parallelization approach, as well as its extension to a distributed memory setting. Finally, we present our conclusions in the context of both LP and the WWW in \S\ref{sec:conclusions}, and motivate our future work in \S\ref{sec:future}.

\section{Link Prediction Background}\label{sec:LP_back}

LP algorithms are often classified in three families \cite{LPCN}: statistical relational models (SRM), maximum likelihood algorithms (MLA) and similarity-based algorithms (SBA). SRM may include probabilistic methods. Probabilistic SRM build a joint probability distribution representing the graph based on its edges, and estimate the likelihood of edges not found in the graph through inference \cite{SRL}. These are frequently based on Markov Networks or Bayesian Networks. Non-probabilistic SRM are often based on tensor factorization, a useful approach for predicting edges in heterogeneous networks (\ie composed by more than one type of relation) \cite{TENSOR}. These have also been used in combination with probabilistic models \cite{PROB_TENSOR}. The second type of LP algorithms, MLA, assume a given structure within the graph (\eg a hierarchy, a set of communities, \etc) and fit a graph model through maximum likelihood estimation. Based on the model, MLA calculate the likelihood of non-existing edges. MLA provide relevant insight into the composition of the graph (\ie how is its topology defined and why), information that can be used for other purposes beyond LP. An example of MLA is the Hierarchical Random Graph \cite{CLA} which builds a dendrogram model representing a hierarchical abstraction of the graph, and obtains the connection probabilities that most accurately represents the graph hierarchically.

The third type of LP methods, SBA, compute a similarity score for each pair of vertices given their topological neighbourhood. Each edge is evaluated on its own, thus potentially in parallel, without previously computing a graph model (unlike SRM and MLA). To score an edge, SBA focus on the paths between the pair of vertices. As a result SBA can be categorized in three classes according to the maximum path length they explore: \emph{local} if they consider only paths of length 2, \emph{global} if they consider paths without length constrain, and \emph{quasi-local} if they consider limited path lengths larger than 2. Global SBA effectively build a model of the whole graph, with the consequent computational cost and poor scalability \cite{LPCN}. Quasi-local SBA are a compromise between between efficacy and efficiency, and are often based on the random walk model \cite{LRW,NEWPERS} or on the number of paths between the pair of vertices \cite{LP}. 

Beyond the previous classification of LP methods, there are more heterogeneous solutions. In \cite{DLP}, the LP process is decomposed in two steps: a first macro-processing step obtains clusters in the graph satisfying certain structural properties, while a second micro-processing step tries to find new links on those clusters based on attribute properties. This work is particularly appropriate for heterogeneous networks that grow over time. The same bi-scale approach is taken in \cite{BISCALE}, where the mesoscale and the microscale components are combined through a statistical inference model. Significantly, the microscale component is based on the Resource Allocation algorithm \cite{RA}.
A different solution is proposed in \cite{NEWPERS}, where authors identify 12 topological features of directed weighted links which are used as input to a set of traditional DM classifiers. To make the approach feasible, link features are obtained at a local level, considering only the sub-graph at 2 to 4 hops away. Results indicate that their proposed algorithm (HPLP), which includes CN as feature, implements an ensemble of classifiers through bagging and random forests, and performs undersampling for correcting the class imbalance, outperforms SBA on both directed and an undirected weighted graphs.

\subsection{Local similarity-based algorithms}\label{sec:local}

Local SBA are among the most scalable approaches to LP, as these methods only need to explore the direct neighbourhood of a given pair of vertices to estimate the likelihood of a edge between them. Since this estimation can be done in parallel (it does not depend on the likelihood of other edges) these methods scale very efficiently and hardly waste any resources. From the perspective of similarity scores, local are the most essential of SBA, as all quasi-local originate from a local one: when the number of steps is set to one, quasi-local SBA are equivalent to some local SBA \cite{KNOWLOD}. This motivates the research on better local SBA from which better quasi-local SBA can be derived. For the remaining of this paper we will focus on SBA methods.

Similarity-based algorithms were first compared in \cite{LKPRE}. Nine 
algorithms were tested on five different scientific co-authorship graphs in the field of physics and compared with a random predictor as baseline. Although no method clearly outperformed the rest in all datasets, three methods consistently achieved the best results; local algorithms Adamic/Adar and Common Neighbours, and global algorithm Katz \cite{KZ}. In \cite{LPSP} the same results were obtained, with Adamic/Adar and Common Neighbours achieving the best results among local algorithms. In \cite{RA} a new local algorithm called Resource Allocation was proposed and compared with other local similarity-based algorithms. Testing on six different datasets showed once again that Adamic/Adar and Common Neighbours provide the best results among local algorithms, but it also showed how Resource Allocation was capable of improving them both. 

The Common Neighbours (CN) algorithm \cite{CN} computes the similarity $s_{x,y}$ between two vertices $x$ and $y$ as the size of the intersection of their neighbours. Formally, let $\Gamma(x)$ be the set of neighbours of $x$ 

\begin{definition}\label{alg:CN}
$$s_{x,y}^{CN}=|\Gamma(x)\cap \Gamma(y)|$$
\end{definition}

The Adamic/Adar (AA) algorithm \cite{AA} follows the same idea as CN, but it also considers the \emph{rareness} of edges. To do so, shared neighbours are weighted by their own degree and the score becomes

\begin{definition}\label{alg:AA}
$$s_{x,y}^{AA}=\sum\limits_{z\in\Gamma(x)\cap \Gamma(y)}\frac{1}{log(|\Gamma(z)|)}$$
\end{definition}

The Resource Allocation (RA) algorithm \cite{RA} is motivated by the resource allocation process of networks. In the algorithm's simpler implementation, each vertex distributes a single resource unit evenly among its neighbours. In this case, the similarity between vertices $x$ and $y$ becomes the amount of resource obtained by $y$ from $x$

\begin{definition}\label{alg:RA}
$$s_{x,y}^{RA}=\sum\limits_{z\in\Gamma(x)\cap \Gamma(y)}\frac{1}{|\Gamma(z)|}$$
\end{definition}

\subsection{Link Prediction for the WWW}\label{sec:lp_www}

The first application that comes into mind when considering the application of LP to the WWW is to discover missing hyperlinks. However, there are few contributions which specifically focus on the application of SBA for such purpose. In contrast, other similar problems like predicting relations in social network graphs, have been thoroughly studied using a wide variety of methodologies \cite{AA,LKPRE,NEWPERS,LPWEIGHT,GLOBALOPTIM}. 

Webgraphs have been occasionally used to evaluate LP scores, but always in combination with other graphs types, and never as an independent case of study. When webgraphs have been used for LP, only relatively small graphs have been computed (a few thousand vertices), typically due to the complexity of the models being used \cite{LP,DIRECTEDRW}. An exception to the lack of attention to hyperlink prediction is the Wikipedia webgraph, composed by Wikipedia articles and the hyperlinks among those. The encyclopedic information contained in the vertices of this graph has served as a motivation to researchers, who have used it to enrich the learning process. As a result most LP methods used on the Wikipedia webgraph are heterogeneous combinations of ML methods that can be neither scaled to large graphs, nor generalized to other webgraphs. In \cite{MISSW} authors propose an algorithm combining clustering, natural language processing and information retrieval for finding missing hyperlinks among Wikipedia pages. With a more specific target in mind, in \cite{MISSW2} authors use supervised learning to train a disambiguation classifier, in order to choose the appropriate target for article hyperlinks. Multiple contributions also exist from the semantic web community, typically using RDF data to implement link inference methods \cite{MISSW3}.

\section{Hierarchies in the WWW}\label{sec:web_hier}

Hierarchies are the most widely used knowledge organization structure. They can be found in domains such as the human brain, metabolic networks, terrorist groups, protein interactions, food webs, social networks, and many others. Within the field of link prediction hierarchies have been acknowledged as a relevant source of information due to its power to describe knowledge \cite{SR}. However, while link prediction methods are predominantly focused on undirected structures for the sake of simplicity, hierarchies are necessarily directed. This structural divergence complicates the use of hierarchies for traditional link prediction, and is the main reason why this combination have not been further exploited so far.

One of the main contributions of our work is the adaptation and application of a LP score based on hierarchies to the problem of hyperlink prediction. By doing so we are implicitly assuming that the WWW topology contains or is partly defined by hierarchical properties. That assumption has been discussed, empirically evaluated and exploited in the past, as presented in this section. What is novel in our contribution is its successful application to the use case of hyperlink discovery. As we will see in \S\ref{sec:experiments}, we consider hierarchies at a much smaller scale than usual, thus enabling novel hierarchical properties and applications in the context of the WWW.

The topology of routing units \cite{HIERAS}, and the WWW requests and traffic \cite{TRAFFHIER} were the first aspects of the Internet shown to have hierarchical properties. The webgraph topology itself was found to compose a self-similar structure defined by a set of nested entities in \cite{WEBHIER}, indicating a \emph{natural hierarchical characterization of the structure of the web\/}. This idea was extended in \cite{HIERNET}, where authors showed how some key properties found in webgraphs and other real networks, such as being scale-free and having a high degree of clustering, could be satisfied through a hierarchical organization. In their work, Ravasz and Barb\'{a}si propose a hierarchical network model fitting such networks, to be composed by small, dense groups, recursively combining to compose larger and sparser groups, in a fractal-like structure.

The importance of hierarchies to model the structure of real networks was explored through its application to generative models: models built to produce artificial, large scale networks mimicking the topological properties of real networks. The work of Leskovec \etal\ \cite{GRAPHSTIME} is of particular interest, as authors define generative models satisfying complex properties never captured before (\eg densification, shrinking diameters and a heavy-tailed out-degree). The most complex model proposed by Leskovec \etal, the forest fire model (FFM), builds a network by iteratively adding one new vertex \emph{x} to a graph. In this model, one randomly provides an entry point to \emph{x} in the form of a vertex \emph{x} will point to through a first out-going edge (\ie its ambassador). In a second step, a random number of out-going edges are added to \emph{x}, to be chosen among the out-going and in-going edges of the \emph{ambassador\/} (the former with a higher probability of being selected than the latter). This process is then repeated recursively for each new vertex that gets connected with \emph{x}, with the restriction that each \emph{ambassador\/} is considered only once. Notice that the FFM methodology contains no explicit hierarchy, and yet it generates data which is hierarchically structured. We will discuss how and why this happens in \S\ref{sec:conclusions}, by comparing FFM with our proposed LP algorithm.

\section{The Hybrid-INF Score}\label{sec:tunn}

The local SBA obtaining the best predictive results so far are all based on the accumulated number of shared neighbours (\eg CN). Additionally, some of these scores weight the evidence provided by each neighbour by its own degree, thus increasing the importance of rare edges (\eg RA, AA) \cite{LKPRE,LPSP,RA}. CN and AA have been shown to achieve the best results among local SBA in citation and social networks \cite{LKPRE,LPSP}, while RA was capable of achieving similar results to those of CN and AA in several domains, and even to improve them in a webgraph of political blogs \cite{RA}. To emphasize the importance of these type of scores, let us remark that the current best quasi-local SBA scores are based on one either CN, RA or AA \cite{KNOWLOD}, and most alternative solutions also include them \cite{DLP,BISCALE,NEWPERS}. This illustrates the potential benefits of obtaining better local SBA.

Given the scarce information available to local SBA (\ie the topological neighbourhood around a pair of vertices) it seems difficult to find scores which are significantly more precise than the current best. As one of the main contributions of this article we argue that in the presence of large amounts of data, implicit graph models may naturally exist, and that these models can be used to significantly improve the performance of LP. A model assumed to exist and exploited for learning has the benefit of providing additional knowledge on the structure of the network, while not adding any complexity (\eg building the model). As discussed in \S\ref{sec:web_hier}, the WWW webgraph is strongly related with hierarchical organizations, which leads us to believe that the hyperlink prediction process could be improved by the assumption of a hierarchical model. 

Hierarchies are at the core of INFerence (INF \cite{KNOWLOD}), a LP score assuming hierarchical properties in the topology to guide the LP process. INF was designed for mining explicitly hierarchical graphs, implementing properties like edge transitivity. The original definition of INF however seems inappropriate for graphs without an explicit hierarchical structure, as INF assumes the computed graph implicitly implements a hierarchy. Based on this assumption INF defines a score for each ordered pair of vertices measuring its hierarchical evidence, providing higher scores to hierarchically coherent edges. So far INF has been shown to be a very good predictor for graphs containing explicitly hierarchical relations, such as the ontology \emph{is-a} relation and the \emph{hyponymy/hyperonymy} linguistic relation of WordNet \cite{KNOWLOD}. Given a directed graph $G=(N,E)$, and a given vertex $x\in N$, the original definition of INF refers to the vertices linked by an out-going edge from \emph{x} as the \emph{ancestors} of \emph{x} ($A(x)$), and to the vertices linked by an in-going edge from \emph{x} as the \emph{descendants} of \emph{x} ($D(x)$).

\begin{definition}\label{def:base}
$$\forall x,y\in N: y \in A(x) \leftrightarrow {x\rightarrow y} \in E$$
$$\forall x,y\in N: y \in D(x) \leftrightarrow {y\rightarrow x} \in E$$
\end{definition}

\begin{figure}
\centerline{\includegraphics[width=0.45\textwidth]{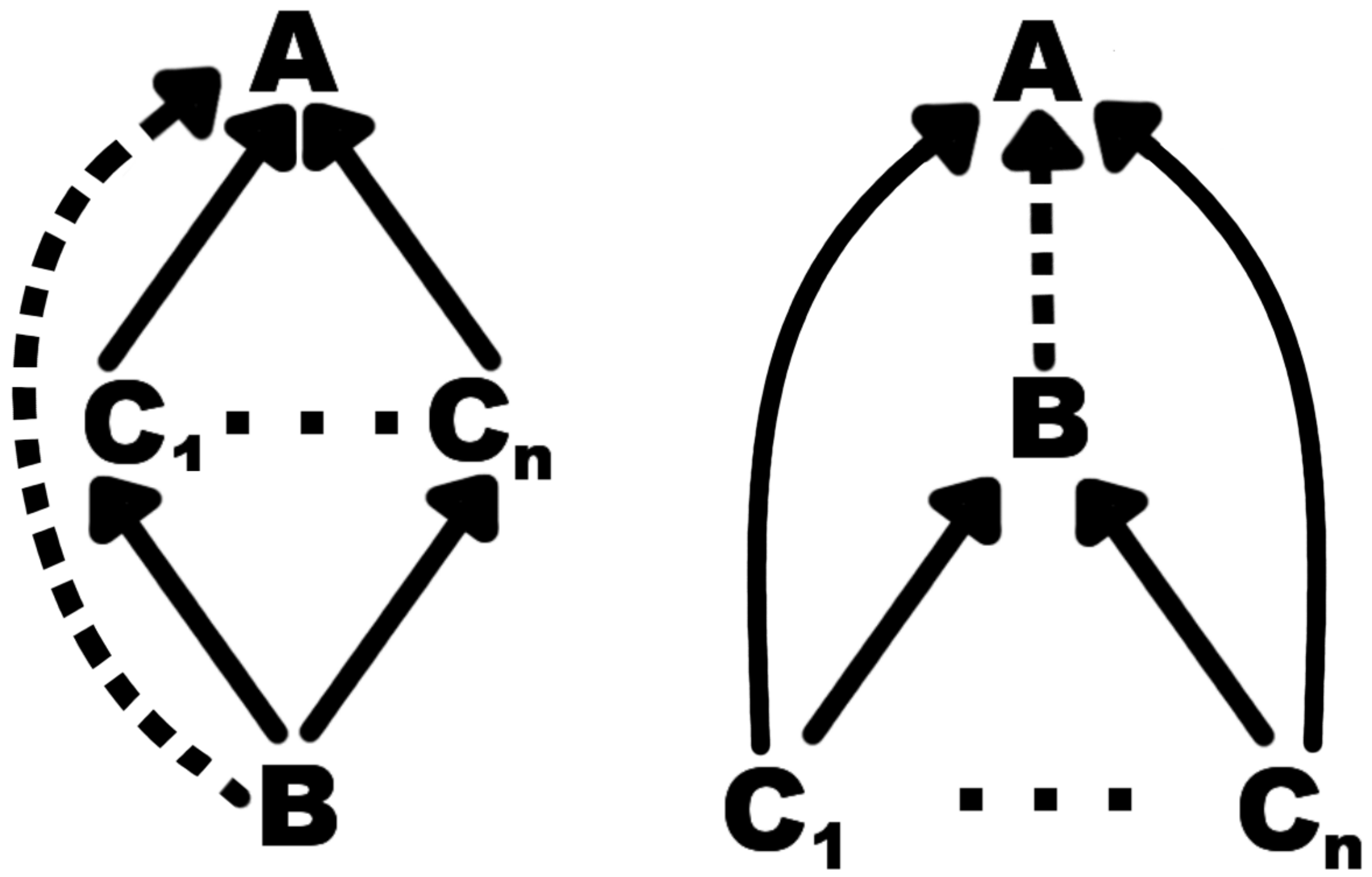}}
\caption{Dashed edge represents link inference. On the left: graphic representation of the top-down DEDuctive process for estimating link likelihood. On the right: graphic representation of the bottom-up INDuctive process for estimating link likelihood. On both cases C nodes represent the evidence considered by each sub-score.}
\label{fig:deduction_induction}
\end{figure}

Based on those sets INF defines two sub-scores named \emph{deductive} sub-score (DED) and \emph{inductive} sub-score (IND), in coherency with a specialization/generalization sense between descendants/ancestors. The DED sub-score follows a top-down reasoning process from the abstract to the specific, resembling that of a weighted deductive inference: \emph{if my four grandparents are mortal, I will probably be mortal too}. In this case information of a vertex ($me\rightarrow mortal$) is obtained from the ancestors of that vertex ($me\rightarrow$ $grandparent$ $\rightarrow$ $mortal$), in proportion with the number of times the relation is satisfied (\eg four out of four if I do not have any living grandparent). Hence, the more of our generalizations share a property, the more certain we can be about that property applying to ourselves. IND on the other hand follows a bottom-up reasoning process from the specific to the generic, resembling that of a weighted inductive inference: \emph{if most of an author's publications are meticulous, the author will most likely be meticulous}. In this example, the information of a vertex ($author\rightarrow meticulous$) is obtained from the descendants of that vertex ($publication\rightarrow meticulous$, $publication\rightarrow author$), in proportion with the number of times the relation is satisfied. Hence, the more frequently our specializations share a property, the more certain we can be about that property applying to ourselves. See Figure \ref{fig:deduction_induction} for a graphical representation of these processes. INF is then defined as $DED+IND$, as in Definition \S\ref{alg:INF}.

\begin{definition}\label{alg:INF}
$$s_{x\rightarrow y}^{INF}=\frac{|A(x) \cap D(y)|}{|A(x)|} + \frac{|D(x) \cap D(y)|}{|D(x)|}$$
\end{definition}

The INF algorithm as defined in Definition \S\ref{alg:INF} maximizes the predictions of links according to a natural sense of hierarchy. Anti-hierarchical evidence, such as the one provided by the common neighbour Z in the setting $X\leftarrow Z \leftarrow Y$, is not considered when evaluating the score of the edge $X\rightarrow Y$. This is potentially beneficial for domains which are implicitly hierarchical, such as webgraphs. However, INF is a purely proportional score, which makes it unsuited for real world data.

\subsection{Proportional and Accumulative Scores}

INF was originally defined to target formal graphs \cite{KNOWLOD} where relations have a high reliability due to an expert validation process (\eg WordNet), or because they originate from formal properties (\eg ontologies). To exploit this fact, INF was entirely based on the proportion of evidence (see Definition \S\ref{alg:INF}) as in formal graphs a single relation is as reliable as many, regardless of size (\eg one out of one vs ten out of ten). Unfortunately, proportional evidence is not equally reliable when working with informal graphs (\eg webgraphs, social networks, costumer-item networks, \etc), where edges often contain errors, outliers or imbalanced data. In this setting, considering a \emph{one out of one} evidence set as a certain scenario would be precipitous and prone to error. 

An alternative to proportional evidence is accumulative evidence, the most frequent approach used by LP algorithms. \emph{Proportional scores}, such as INF or Jaccard's coefficient \cite{LKPRE}, weight the evidence of edges according to their local context, and provide a normalized similarity for each edge regardless of their degree. This makes proportional scores unbiased towards edges among high-degree vertices. \emph{Accumulative scores} on the other hand measure the absolute amount of evidence, ignoring the local context (\eg source vertex degree). In these scores, edges are evaluated and ranked from a graph-wide perspective, which benefits predictions around high-degree vertices.


Previous results have shown that in general, accumulative scores perform better than proportional scores: the top three scores found in the bibliography are accumulative (\ie CN, AA and RA). This can be explained by the importance of the preferential attachment process (\ie the rich get richer), which accumulative scores implicitly satisfy, and proportional scores avoid. But the generally poor results of proportional scores is also caused by the volatility of predictions among low-degree vertices, which proportional scores do not neglect, and which are often less reliable due to arbitrariness.

\subsection{From INF to a Hybrid Score}\label{sec:INF}

To adapt the INF score to informal domains such as webgraphs, we extend it to consider both proportional and accumulative evidence. Although most current solutions are purely accumulative, we build a hybrid solution with the goal of combining the certainty of proportional scores with the reliability of accumulative ones. To do so we first normalize evidence given the local context (proportionally, through a division) and then weight that proportion through the absolute size it was based on (accumulatively, through a logarithm). By using a logarithmic function to weight the score, the accumulative evidence dominates the scores of low degree vertices, while the proportional evidence remains more important for high degree vertices. The $INF\_LOG$ variant is defined in Definition \S\ref{alg:INF_LOG}.

\begin{definition}\label{alg:INF_LOG}
$$s_{x\rightarrow y}^{INF\_LOG}=\frac{|A(x) \cap D(y)|}{|A(x)|}*log(|A(x)|) + \frac{|D(x) \cap D(y)|}{|D(x)|}*log(|D(x)|)$$
\end{definition}

The original INF definitions combines the evidence provided by DED and IND through an addition ($INF=DED+IND$), considering both of them equally reliable. However, through a preliminary evaluation we found that DED consistently achieves higher precisions than IND on most domains. Our hypothesis to understand this behaviour is two-fold. For formal graphs, the deductive reasoning process in which DED is based may be more reliable than the inductive reasoning process of IND. While for informal graphs, this behaviour may be caused by the fact that a vertex typically has a higher responsibility on defining the set of vertices it points to, while the set of vertices that point to it are often imposed onto it (\eg webmasters rarely choose who links to their site). A higher reliability of the outgoing edges of a vertex would make DED more precise at defining the vertex itself, as those are the edges used by DED. 

Even though DED is more reliable than IND, the contribution of IND to the INF score is still relevant, as it is capable of detecting a type of relational evidence DED does not consider. We solve this problem by adding a multiplying factor to DED ($INF\_kD=k*DED+IND$), as seen in Definition \S\ref{alg:INF_LOG_kD}

\begin{definition}\label{alg:INF_LOG_kD}
$$s_{x\rightarrow y}^{INF\_LOG\_kD}=k*\frac{|A(x) \cap D(y)|}{|A(x)|}*log(|A(x)|) + \frac{|D(x) \cap D(y)|}{|D(x)|}*log(|D(x)|)$$
\end{definition}

We sampled the $k$ value for several domains, and found the optimal value to be between 1 and 3 for most graphs. To provide a consistent score evaluation in \S\ref{sec:experiments}, we set $k=2$ in all our tests. This implementation of the Hybrid-INF score is identified as $INF\_LOG\_2D$.

%

\section{Experiments}\label{sec:experiments}

Next we evaluate the INF\_LOG\_2D algorithm for the hyperlink prediction problem within its same family of algorithms (local SBA scores). Although the hyperlink prediction problem is a directed link prediction problem, we use CN, AA and RA (all undirected link prediction scores) as baseline because these are the current state-of-the-art in LP, and because there are no analogous directed local SBA scores in the bibliography. We compare the performance of INF\_LOG\_2D, CN, AA and RA, on six webgraphs obtained from different sources: the webgraph of the University of Notre Dame domain in 1999 (webND) \cite{WEBND}, the webgraph of Stanford and Stanford-Berkley in 2002 (webSB) \cite{WEBSB}, a webgraph provided by Google for its 2002 programming contest (webGL), a domain level webgraph of the Erd\H{o}s WebGraph project from 2015 (erd\H{o}s), and the webgraphs of the Chinese encyclopedias Baidu and Hudong \cite{KONECT}. Sizes of all graphs are shown in Table \ref{tab:sizes}. All graphs used are publicly available through the referenced sources.

\begin{table}[h]
 \centering
  \begin{tabular}{| c | r | r | }
    \hline
    \bf{Graph} & \bf{Vertices} & \bf{Edges} \\ \hline
      webND 	& 325,729	& 1,497,134\\ \hline
      webSB 	& 685,230	& 7,600,595\\ \hline
      webGL 	& 875,713	& 5,105,039\\ \hline
      erd\H{o}s& 1,817,390	& 16,391,889\\ \hline
      hudong 	& 1,984,484 	& 14,869,484\\ \hline
      baidu 	& 2,141,300 	& 17,794,839\\ \hline
  \end{tabular}
\caption{Size of computed webgraphs.}
\label{tab:sizes}
\end{table}

For its evaluation, the LP problem is reduced to a binary classification problem. In this context one has a set of correct instances (edges missing from the graph known to be correct) and a set of wrong instances (the rest of missing edges) and the goal of the predictors is to classify both sets of edges as well as possible. The most frequently used metrics to evaluate the predictors performance in this context are the Receiver Operating Characteristic (ROC) curve and the Precision-Recall (PR) curve.

The ROC curve sets the True Positive Rate (TPR) against the False Positive Rate (FPR), making this metric unbiased towards entities of any class regardless of the size of classes. Unfortunately, their consideration of errors can result in mistakenly optimistic interpretations \cite{ROC_PR,EVALLP}, as ROC curves represent miss-classifications relative to the number of errors that \emph{could be made}. In domains where the negative class is very large and one can make millions or even billions of errors, showing mistakes as relative to the negative class size (\ie FPR) may hide their actual magnitude and complicate a realistic assessment of predictive performance. Furthermore, in large and highly imbalanced domains most of the ROC curve becomes irrelevant in practice, as it represents inapplicable precisions below 1\% \cite{CCIA15}. 

Precision-recall (PR) curves are an alternative to ROC curves, which show precision on the \emph{y} axis and recall on the \emph{x} axis. PR curves do not show the number of correct classifications for the negative class, and instead represent miss-classifications relative to the number of predictions made. This allows for a straight-forward idea of the actual predictive quality in absolute terms, and makes the whole curve relevant regardless of the problem size. In fact, ROC and PR curves are strongly related, as a curve dominates another (it is above it) in the ROC space if and only if it also dominates it in PR space \cite{ROC_PR}. The main difference between ROC and PR curves is on how errors are represented, but this difference has a huge visual impact. Consider for example how the two curves represent a random classifier, which always performs poorly in a large and highly imbalanced data set. The ROC curve always represents the random classifier as a straight line between points $(0,0)$ and $(1,1)$, regardless of class imbalance, with all better than random classifiers represented as lines above that diagonal. The more demanding PR curves on the other hand represent random classifiers in imbalanced data sets a flat line on the \emph{x} axis, as their precision in imbalanced settings is always close to zero. 

Due to the previously outlined motives and the huge class imbalance found in our graphs (see Table \ref{tab:problem_size}), we will use the Area Under the Curve measure of the PR curve (AUPR) to compare the various scores evaluated. Let us remark this is the recommended approach in this setting \cite{EVALLP}, even though it has not been fully assimilated by the community yet.




To calculate the PR curve one needs a source graph (on which we wish to predict edges) and a test set (edges missing from the source graph known to be correct). To build the test set we randomly split the edges of each graph 90\%-10\%, and use the 90\% as source graph and the remaining 10\% as test set. The size of test sets for each evaluated graph are shown in Table \ref{tab:problem_size}. Vertices becoming disconnected in the source graph after removing the test edges were not computed nor considered for score evaluation. 

A frequently used methodology to build test sets is 10-fold cross validation. However, due to the size of the graphs being used this is not necessary. The law of large numbers will make any significant portion (\eg 10\%) of a large domain tend towards a stable sample, thus making a single run a representative and accurate sample of the performance \cite{CCIA15}.

We build all PR curves exhaustively, as all possible edges in the graph are computed. Approximate and potentially dishonest methodologies such as test sampling \cite{EVALLP} were avoided. Computing all possible edges for graphs the size of the ones used here equals to the evaluation of billions of edges (see Table \ref{tab:problem_size}), and any proposed solution needs to be scalable in order to be feasible. SBA fit this requirement perfectly, as they can be parallelized with maximum computational efficiency. See \S\ref{sec:large} for details on the implementation and parallelization of the solution.

\begin{table}
 \centering
  \begin{tabular}{| c | r | r | l |}
    \hline
    \bf{Webgraph} 	& \bf{Positive edges} 	& \bf{Negative edges} 	& \bf{Class imbalance} \\ \hline
      webND 	& 133,279		& 95,939 million 	& 1:719,835\\ \hline
      webSB 	& 756,937		& 466,034 million 	& 1:615,684\\ \hline
      webGL 	& 494,982		& 741,978 million 	& 1:1,498,999\\ \hline
      erd\H{o}s	& 1,543,090		& 2,970,460 million 	& 1:1,925,007\\ \hline
      hudong 	& 1,446,760		& 3,786,991 million 	& 1:2,617,566\\ \hline
      baidu 	& 1,701,330		& 4,370,982 million 	& 1:2,569,155\\ \hline
      
  \end{tabular}
\caption{For each graph, number of correct edges and number of incorrect edges being computed, and positive:negative class imbalance in the evaluated set.}
\label{tab:problem_size}
\end{table}

\subsection{Results Analysis}\label{sec:results}

\begin{figure}
\centerline{\includegraphics[scale=0.12]{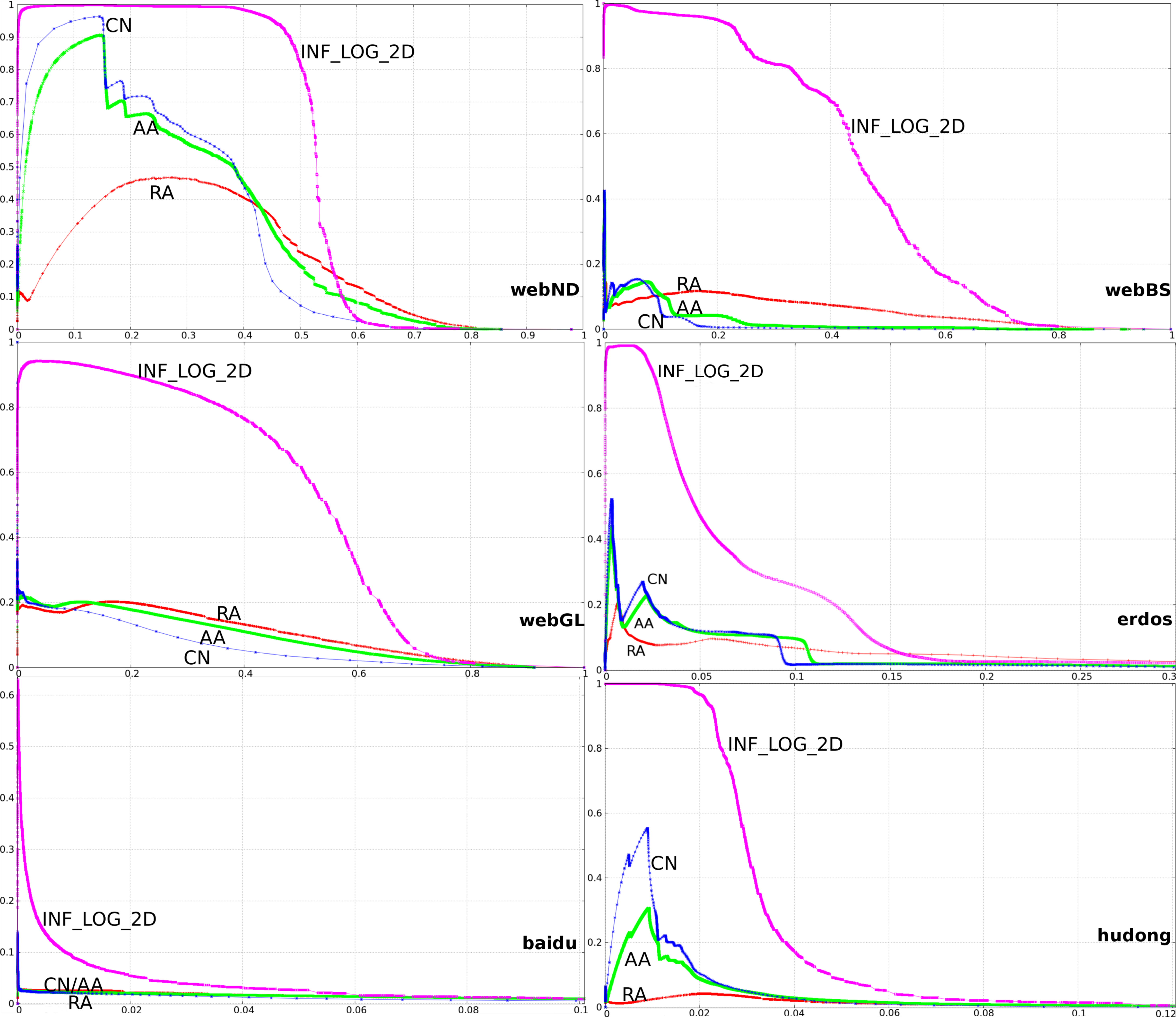}}
\caption{PR curves for all webgraphs. Recall in \emph{x} axis, precision in \emph{y} axis. Erd\H{o}s, hudong and, baidu curves are zoomed in for clarity.}
\label{fig:curves}
\end{figure}

The PR curves of all four scores on all six graphs can be seen in Figure \ref{fig:curves}. INF\_LOG\_2D achieves the best predictive performance on all graphs by a large margin. The huge improvement in precision obtained by INF\_LOG\_2D (\ie much higher $y$ axis values on its curve), results in an increase of up to one order of magnitude in the AUPR over the current state-of-the-art (see Table \ref{tab:aucs}). Significantly, INF\_LOG\_2D achieves precisions of 100\% at a very small recall, even when the other scores never reach 50\% precisions. As seen in these results, INF\_LOG\_2D can recommend thousands of edges (those with higher reliability) making very few mistakes in the process. The leap in performance of INF\_LOG\_2D, and its consistence within webgraphs, also stresses the importance of hierarchical properties for hyperlink prediction, and how much it can be gained by integrating implicit network models within predictive algorithms.

\begin{table}
 \centering
  \begin{tabular}{| c | c | c | c| c | r | }
    \hline
			& \bf{AA} 	& \bf{CN}	& \bf{RA}  	& \bf{INF\_LOG\_2D} 	&\bf{Improvement}\\ \hline\hline
      \bf{webND}	& 0.31679	& 0.31855	& 0.21178	& 0.52640	& 65.24\% (CN)\\ \hline\hline
      \bf{webSB} 	& 0.02218	& 0.01669	& 0.05491	& 0.45156	& 722.36\% (RA)\\ \hline\hline
      \bf{webGL} 	& 0.08961	& 0.06227	& 0.10035	& 0.49210	& 390.38\% (RA)\\ \hline\hline
      \bf{erd\H{o}s} 	& 0.01933  	& 0.01826	& 0.02220	& 0.07299	& 228.78\% (RA)\\ \hline
      \bf{hudong} 	& 0.00555	& 0.00743	& 0.00223	& 0.03424	& 360.83\% (CN)\\ \hline\hline
      \bf{baidu} 	& 0.00285	& 0.00176	& 0.00308	& 0.00528	& 71.42\% (RA)\\ \hline
  \end{tabular}
\caption{AUC obtained by each score on the PR curves shown in Figure \ref{fig:curves}. Also, improvement over best accumulative score in percentage of PR-AUC.}
\label{tab:aucs}
\end{table}


\subsection{Hybrid Improvement}\label{sec:over_INF}

INF\_LOG\_2D is a hybrid score which outperforms the best accumulative scores on different six webgraphs (see Table \ref{tab:aucs}). So far, purely accumulative scores had obtained the best results on almost every informal graph evaluated, which makes these results all the more important. To validate the importance of the hybrid approach also when compared to purely proportional scores, we tested a well-known proportional score on the same six webgraphs, the \emph{Jaccard's coefficient} score \cite{LKPRE}. Where $\Gamma(x)$ is the set of neighbours of $x$, Jaccard's coefficient is defined as

\begin{definition}\label{alg:JACC}
$$s_{x,y}^{Jaccard}=\frac{|\Gamma(x)\cap \Gamma(y)|}{|\Gamma(x)\cup \Gamma(y)|}$$
\end{definition}

Our results showed that Jaccard's performance was incomparably worse on all webgraphs; when plotted together with the other scores, Jaccard's PR curve was a flat line on the \emph{x} axis on all graphs. Jaccard's is not shown in Tables \ref{tab:aucs} or \S\ref{tab:over-inf} because its AUPR was always zero given the five decimals used. Consistently with previous research, we find proportional scores to be very imprecise on their own. However, for the first time we show that proportional scores can help improve significantly the more precise accumulative scores when integrated.

\begin{table}
 \centering
\setlength{\tabcolsep}{0.4em}
  \begin{tabular}{| c | c | c | c | c | c | c |}
    \hline
    \bf{Webgraph} 		&  webND	&webSB		&webGL		&erd\H{o}s	&hudong		&baidu\\ \hline
    \bf{INF PR-AUC}		&0.09966	&0.10530	&0.12826	&0.00698	&0.00402	&0.00065\\ \hline
    \bf{Accum. PR-AUC}		&0.31855	&0.05491	&0.10035	&0.02220	&0.00743	&0.00308\\ \hline
    \bf{Hybrid Improve.} 	&  428\%	& 328\% 	&283\% 		& 945\%		& 751\%		& 712\%\\ \hline
      
  \end{tabular}
\caption{For each graph tested, PR-AUC of the basic INF score, PR-AUC of the top accumulative score, and percentage of PR-AUC improvement of INF\_LOG\_2D over INF}
\label{tab:over-inf}
\end{table}

To explore the impact of making hierarchical assumptions on webgraphs, we now consider the performance of the basic INF score. INF is a purely proportional score assuming a hierarchical model in the graph, and it actually outperforms the best accumulative scores on two of the six webgraphs (see Table \S\ref{tab:over-inf}). INF is incomparably better than the also proportional Jaccard's coefficient, which shows how the handicaps of using a proportional approach can be overcome by exploiting implicit data models found in the graph. 

Finally, let us compare the performance of INF\_LOG\_2D with that of INF. As defined in \S\ref{sec:INF}, INF\_LOG\_2D is a hybrid version of the purely proportional score INF. Their comparison is therefore a reliable test on the benefits of turning a proportional score into a hybrid score. In that regard, Table \ref{tab:over-inf} shows that INF\_LOG\_2D consistently improves the performance of INF by two orders of magnitude in the AUPR measure for all graphs, regardless of INF achieving competitive results on it or not.

\section{Applications}\label{sec:app}

On large graphs, LP is frequently imprecise, does not scale or cannot be generalized. Three reasons which have constrained its application. SBA are clearly generalizable, as they can be applied to virtually any graph composed by vertices and directed edges\footnote{Depending on the score. CN, AA and RA do not require edge directionality. INF\_LOG\_2D does.}. SBA are clearly scalable, as they can be perfectly parallelized to compute graphs of any size. Unfortunately, current SBA are rather imprecise, hardly reaching 50\% precisions on graphs close to 1 million vertices (see Figure \ref{fig:curves}). 

\begin{table}[h]
 \centering
  \begin{tabular}{| c | r | r | r |}
    \hline
			& \bf{AA} 	& \bf{CN}	& \bf{RA}  	\\ \hline
      \bf{webND}	& +66\%		& +65\%		& +148\%	\\ \hline
      \bf{webSB} 	& +1935\%	& +2605\%	& +722\%	\\ \hline
      \bf{webGL} 	& +449\%	& +690\%	& +390\%	\\ \hline
      \bf{erd\H{o}s} 	& +277\%	& +299\%	& +228\%	\\ \hline
      \bf{hudong} 	& +516\%	& +360\%	& +1435\%	\\ \hline
      \bf{baidu} 	& +85\%		& +200\%	& +71\%		\\ \hline

  \end{tabular}
\caption{Percentage of PR-AUC improvement achieved by INF\_LOG\_2D.}
\label{tab:percent}
\end{table}

In this context we have shown how, by considering inherent topological features (\eg hierarchies) and by combining different approaches (\eg proportional and accumulative into hybrid scores), one can overcome the limitation of imprecision. The results obtained by INF\_LOG\_2D are a huge improvement over the previous state-of-the-art (see Table \ref{tab:percent} for details). But more importantly, these results show how SBA can reach precisions between 1 and 0.9, enabling the reliable discovery of tens of thousands of hyperlinks in a straightforward fashion (\ie without building a model of the graph). From these unprecedented results many applications can be derived.



\begin{itemize}
 \item \emph{Web search engines}. Current search engines are composed by a wide variety of interacting metrics, which together produce a complete ranking of web page relevance. The measure of hierarchical similarity between webpages provided by INF\_LOG\_2D may represent a different sort of evidence, and could be used to enrich the ranking of web pages from a different perspective, once the utility of the INF\_LOG\_2D score to characterize webpages within a webgraph has been validated here. Scores such as INF\_LOG\_2D could be combined with algorithms like Page Rank by spreading relevance not only to those webs directly connected, but also to those that LP algorithms estimate as potential neighbours with high reliability.

 \item \emph{Web connectivity}. Hierarchical properties of the WWW spontaneously emerge at a global level (see \S\ref{sec:web_hier}), but at a local level things are rather chaotic. Each webmaster must find appropriate webpages to link to, in a domain with billions of websites. As humans cannot be aware of every single webpage online, a hyperlink recommender could be useful for web masters to find relevant web pages to link to. This tool could improve both the connectivity and coherency of the WWW, or of a specific web domain, as well as significantly enrich directory webs.

 \item \emph{Bottom-up taxonomy building}. Taxonomies are frequently used in online shops and encyclopedias (among others) to organize content. These taxonomies are often defined by external experts, requiring a continuous and expensive manual post-process of data mapping (\eg fitting web pages and articles to taxonomy entities). According to our results it seems feasible to develop an automatic taxonomy building system, proposing a taxonomy of web pages based on their interrelations, similarly to what was proposed by Clauset \etal \cite{CLA}. Such a taxonomy would have the benefit of originating from the data, making it necessarily relevant for the domain in question. It would also be easily updated. A taxonomy like this can be used to optimize the commercial organization of an online shop, for example by considering user navigation paths as source for the LP algorithms.
\end{itemize}


\section{Implementation}\label{sec:large}

In the tests performed in this paper, the LP algorithms compute billions of edges for each processed graph (see Table \ref{tab:times}). Computing these edges sequentially, one by one, is clearly impractical, which makes High Performance Computing (HPC) parallelism necessary for the feasibility of our work. In \S\ref{sec:parallel} we review the algorithmic design and code parallelization we developed to maximize efficiency. In \S\ref{sec:computation} we consider the different computational contexts in which large scale graph mining problems can be set, and how we tackle this particular problem on each of them.

\begin{table}
 \centering
\setlength{\tabcolsep}{0.4em}
  \begin{tabular}{| c | r | r | r | c | }
    \hline
    \bf{Graph} & \bf{Vertices} & \bf{Edges} & \bf{Edges to evaluate} & \bf{Computation time}   \\ \hline
      webND 	& 325,729	& 1,497,134& 9,593.78M &20 seconds\\ \hline
      webSB 	& 685,230	& 7,600,595& 466,028M&21 minutes\\ \hline
      webGL 	& 875,713	& 5,105,039& 741,973M&50 seconds\\ \hline
      erd\H{os}s& 1,817,390	& 16,391,889& 2,970,460M&92 minutes\\ \hline
      hudong 	& 1,984,484 	& 14,869,484& 3,786,970M&49 minutes\\ \hline
      baidu 	& 2,141,300 	& 17,794,839& 4,370,980M&71 minutes\\ \hline
  \end{tabular}
\caption{Size of computed webgraphs, missing edges to be evaluated and time spent evaluating them. Computational context is defined in \S\ref{sec:computation}}
\label{tab:times}
\end{table}

\subsection{Parallelization}\label{sec:parallel}

Large-scale graphs and their need for parallel computing reinforce the importance of SBA, as these are extremely efficient parallel algorithms. The main challenge in the implementation of a graph processing algorithm lies in the high-dimensionality of graphs, which easily translates into \emph{data dependencies}. Dependencies determine execution order constrains among portions of code and imply synchronization points, as one portion of code must wait for another portion to be executed first. Through the existence of dependencies, threads see their work flow halted as they wait for other threads. In essence, dependencies define bottlenecks in the parallel execution of code, and reduce the efficiency of computational resources usage.

A related concept within the field of parallel computing is that of \emph{embarrassingly parallel} problems \cite{EMBPAR}. This notion applies to algorithms that can be parallelized without the definition of significant dependencies. These are therefore problems which can achieve a huge efficiency through parallelization, as there will be almost no idle resources in their computation. Embarrassingly parallel problems are capable of decreasing computational time almost linearly with the number of computing units available, as the various threads must not endure waiting times. As said in \S\ref{sec:LP_back}, one of the key features of similarity based LP algorithms is that the score of each edge can be calculated independently from the rest. This particularity gains a huge relevance now as it allow us to define LP as an embarrassingly parallel problem. Fully testing a LP algorithm on a graph equals to calculate the similarity of all possible ordered pairs of vertices (\ie of each possible directed edge). Since each similarity can be calculated independently, we can evaluate them all simultaneously without dependencies. Considering the huge number of edges to test (at times in the order of billions), the code parallelization design defines the efficiency of the algorithm, and eventually, the size of the graph it can process.

Our algorithmic design is divided into two parallel sections. On the first one we calculate the similarity of each possible edge in parallel, storing the results obtained for the edges originating on each vertex separately (in the \texttt{n1\_partial\_scores} data structure). However, the score of each edge is not computed at once. Instead, evidence is accumulated as we find paths (\texttt{n2} vertices) leading from a given target (\texttt{n1} vertices) to a given goal (\texttt{n3} vertices). An overview of this code can be seen in Algorithm \ref{code:lp1}. Notice that the iterations of the outermost loop (Line 3) can be computed in parallel without dependencies, with the only exception of the storage of results. 

\begin{algorithm}[t]
\SetAlgoNoLine
\KwIn{A graph G.}
\KwOut{Link prediction scores for all vertices in G, stored in a map structure.}
//Map to store all scores $<sourceId,<targetId,score>>$\\
$map<int,map<int,float> >$ $graph\_scores$\;
\For{vertex $n1$ in $G$}{
  //Map to store partial scores $<targetId,partial\_score>$\\
  $map<int,float>$ $n1\_partial\_scores$\;
  \For{vertex $n2$ neighbor of $n1$}{
    \For{vertex $n3$ neighbor of $n2$}{
      //If new target, initialize partial score\\
      \If{$!n3$ in $n1\_partial\_scores$}{
	$n1\_partial\_scores.insert(n3,initial\_score)$\;
      }
      //Else, update partial score\\
      \Else{
      $n1\_partial\_scores(n3).update$\;
      }
    }	
  }
  //Score of edge $n1\rightarrow x$ has been computed for all $x$ in $G$\\
  $graph\_scores.push\_back(n1\_partial\_scores)$\;
}
\caption{Code skeleton for similarity evaluation of all edges in a graph}
\label{code:lp1}
\end{algorithm}

Between the first and second part of the code we perform a reduction, combining the results obtained by all vertices. Since we are interested in evaluating the graph-wide performance, we need to know the number of true positive and false positive predictions achieved at every distinct threshold throughout the graph. This will allow us to simplify the second part of the code, and define it without dependencies.

The second part of the code calculates the performance of LP algorithms. Given the total number of true positive and false positive predictions found at each distinct similarity value, we can calculate the points composing the curves discussed in \S\ref{sec:experiments} through an aggregation process. This task is also embarrassingly parallel, as the performance at each threshold (\ie each point within the PR curves) can be calculated independently from the rest of thresholds. It is important to parallelize this task, as the number of distinct similarity values in large graphs can be also large (up to millions of values). An overview of this second section of code can be seen in Algorithm \ref{code:lp2}.

\begin{algorithm}[t]
\SetAlgoNoLine
\KwIn{The list of all distinct similarities found in a graph, together with their corresponding true and false positive predictions.}
\KwOut{List of performance rates, one per distinct similarity value}
//Structure to store the results obtained at all thresholds\\
$vector$ $<pair<int,int>>$ $full\_results$\;
\For{similarity value $sim1$ in graph}{
  int $true\_pos\_sim1 = 0$\;
  int $false\_pos\_sim1 = 0$\;
  \For{similarity value $sim2$ in graph}{
      \If{$sim2>=sim1$}{
	$true\_pos\_sim1 += true\_pos\_sim2$\;
	$false\_pos\_sim1 += false\_pos\_sim2$\;
      }
  }
  $full\_results.push\_back(sim1,true\_pos\_sim1,false\_pos\_sim1)$\;
}
\caption{Code skeleton for full graph performance evaluation}
\label{code:lp2}
\end{algorithm}

Notice how the outermost loop can be parallelized with the only dependency of writing the results.

\subsection{Computational Setting}\label{sec:computation}

In this paper we computed graphs up to two million vertices. Graphs of this size are already challenging to process exhaustively due to the total number of potential edges in the graph. Nevertheless, LP and the hyperlink prediction problem should target much larger graphs. Recent research is moving in that direction, as shows the work of \cite{NOMAD} capable of training a factorization model for a graph with 50 million vertices. Regardless, the size of some of the most interesting graphs to process remains several orders of magnitude larger than that. For example, a webgraph from 2014 \cite{WEBGRAPH} covered 1.7 billion web pages connected by 64 billion hyperlinks. Large-scale graphs such as these represent the kind of problem for which SBA is the only feasible solution so far.

\begin{figure}
\centerline{\includegraphics[width=0.75\textwidth]{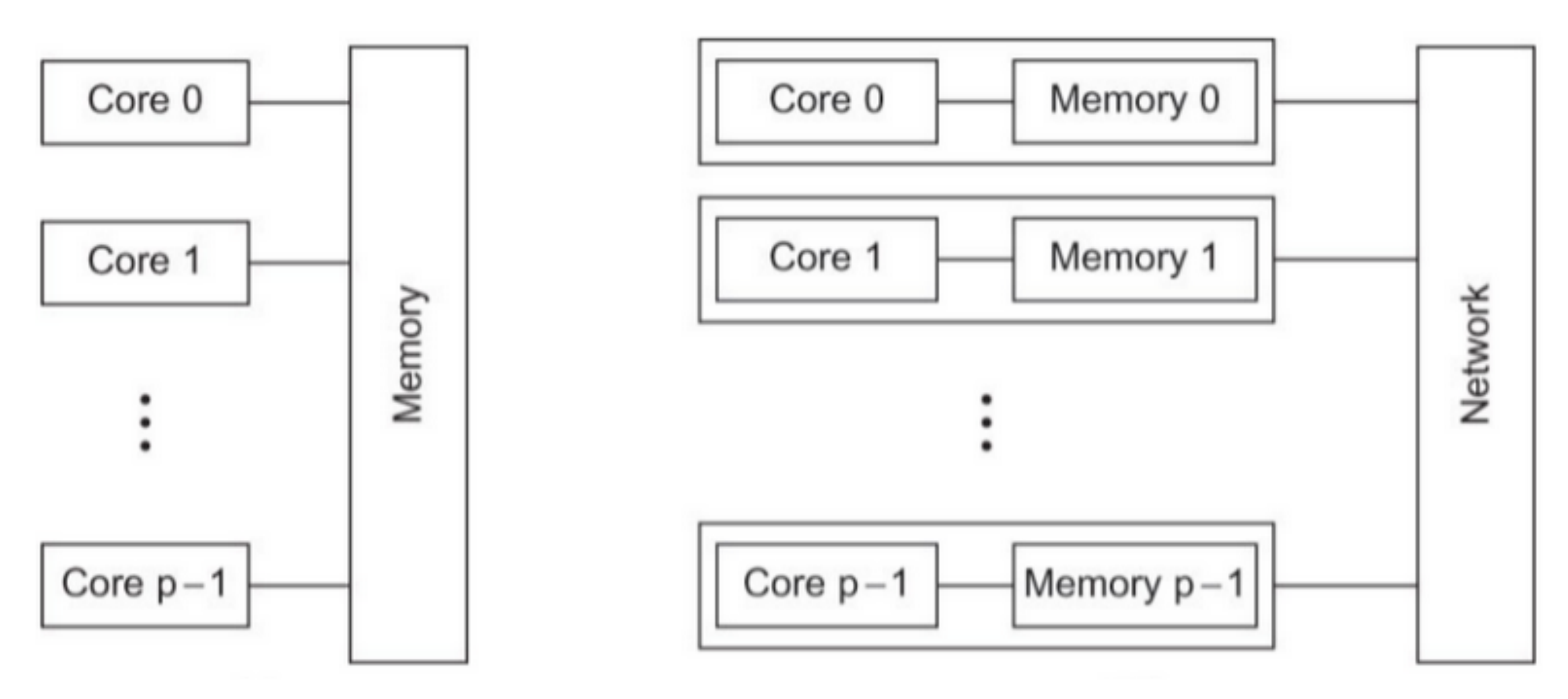}}
\caption{On the left, structure of a shared memory architecture. On the right, structure of a distributed memory architecture.}
\label{fig:memory}
\end{figure}

A main concern when targeting large-scale graphs are computational resources. For medium sized graphs, such as the ones in \S\ref{sec:results}, one can use a \emph{shared memory} context, where all computational units have direct access to a centralized memory space. This approach assumes that the graph data can be stored into a single memory location. An assumption that can hardly be satisfied as graphs grow. Furthermore, even if we manage to store a large-scale graph into a single memory location, the number of computing units (\ie cores) available for parallel execution will be physically constrained. To solve this limitation, HPC researchers use a \emph{distributed memory} setting, where memory is split among several locations. In distributed memory, each memory location is directly accessible only by a subset of all computing units, but data from other locations can be fetched when necessary through communication channels. Using a distributed memory context entails several additional problems, such as how to distribute work among computational resources while achieving balance, and how to split input data among locations as to minimize communications. Problems that remain open issues, but that need to be addressed if we want to process large-scale graphs. See Figure \ref{fig:memory} for a graphical representation of both paradigms.

The code used for the tests presented in this paper was parallelized using the OpenMP shared memory model API \cite{OpenMP}. This API provides a set of compiler directives that extend the C/C++ and FORTRAN base languages. OpenMP directives define how code is to be parallelized and how data is to be shared. We chose OpenMP because it is portable, scalable, flexible and the de-facto standard. Within the first section of code, the parallelization was done on the most external loop (Line 3 of Listing \ref{code:lp1}), effectively distributing its iterations among different threads. This design guarantees that all similarities $n\rightarrow X$ of a given vertex $n$ (\ie a full iteration of the outermost loop) are calculated by a single thread, thus avoiding any dependencies. The second section of the code was also parallelized on the most external loop (Line 3 of Listing \ref{code:lp2}). This design guarantees that each possible threshold (\ie each point within the PR and ROC curves) was calculated by a unique thread, avoiding any dependencies.

When parallelizing a loop through OpenMP one must decide how to distribute iterations among the team of running threads. Of the different ways of splitting iterations, we found that the most efficient for our problem was a \emph{dynamic scheduling}, splitting iterations in chunks of pre-determined size which get assigned to threads as these request it. It is key to define chunk sizes according to the problem size in order to minimize imbalances. If the chunk size is too large, at the end of the computation one or more threads will remain idle for a long time while the rest of threads finish their last chunks. If chunks are too small the constant scheduling of threads, as these finish chunks and requesting for more, may slow down the whole process. In the case of LP, the larger and denser the graph, the smaller the chunks must be, as iterations in a large graph will be more time consuming, thus increasing the possibilities of imbalance. For our code we found the optimal chunk size to be between 100 and 2000.

For our tests we used the MareNostrum supercomputer, provided by BSC. We used a single Intel SandyBridge-EP E5-2670/1600 8-core at 2.6 GHz, with a limit of 28.1 GB of RAM. This translates into 8 parallel threads. The time spent for computing each graph is at most one hour and a half, as shown in Table \ref{tab:times}. Those times include computing the four scores (CN, RA, AA and INF\_LOG\_2D) for all the missing edges of each graph. Time does not include graph loading time, curve building time, and writing of results.

\subsubsection{Distributed Memory}

Local similarity-based scores like the ones we evaluated here can compute graphs with a few million vertices in a shared memory environment. Eventually though, we will be interested in working with larger graphs for which a distributed memory is needed even by local methods. Consider for example the webgraph defined by Internet with 3.5 billion web pages, or a brain connectome graph composed by billions of neurons. For this kind of data sets the only feasible solution nowadays is distributed memory. And not only because of space requirement, but also because of the time complexity. Hundreds of cores computing in parallel will be needed to mine those graphs, and the number of computing cores accessing a single shared memory space is rarely over a few dozens. To run LP methods on a distributed memory environment using the same algorithmic design and parallelization described here we can use the OmpSs programming model \cite{ompss}, which supports OpenMP like directives and has been adapted to work on clusters with distributed memory \cite{bueno}. Even though in the OmpSs version the graph data has to be communicated among computing entities (as the graph data is distributed among locations), our preliminary results show no relevant overhead added by this communication. This is so because in our LP algorithmic design it is easy to predict which edge will be evaluated next, and therefore which graph data will be needed next by each thread (\ie which vertices and neighbours must be brought to memory). Thanks to this foreseeability, data can then sent before it is needed, thus avoiding idle threads and the consequent communication overhead. From a computational point of view this means that LP can scale almost linearly on distributed memory contexts.

\section{Conclusions}\label{sec:conclusions}

We have introduced a novel method which assumes the existence of hierarchical properties to improve the task of hyperlink prediction.
Our first conclusion is that, according to the results shown in \S\ref{sec:results}, the task of hyperlink prediction can be significantly improved through the consideration of hierarchical properties. In our tests the INF\_LOG\_2D hierarchical score outperformed all non-hierarchical scores (CN, RA and AA) on six different webgraphs, doubling its AUPR measure in the worse case. The size and variety of the graphs used, and the thoroughness of the evaluation methodology guarantee the consistency of these results. Results that align with the previous work discussed in \S\ref{sec:web_hier}, providing further evidence on the importance of hierarchies for defining the topology of webgraphs.

From a practical point of view, our main conclusion is that, through the leap in performance obtained shown here we can now predict thousands of hyperlinks in a webgraph with almost perfect precision, and in a scalable manner. As seen in Figure \ref{fig:curves}, INF\_LOG\_2D achieves precisions close to 100\% through its most certain predictions. According to our results, by considering hierarchies hyperlink prediction becomes a feasible problem. This immediately enables multiple interesting cases of application, including but not limited to: increase and improve the connectivity of web pages, optimize the navigability of web sites, tune search engines results through web similarity analysis, and refine product recommendation through item page linkage.

From a general WWW perspective, this work continues the analysis of the relation between hierarchies and the WWW. A step in a different direction than most contributions, typically focused on defining generative models of real networks. Generative models and the problem of LP are strongly related, since generative models must produce new edges within a given graph. To illustrate that let us consider the FFM, briefly described in \S\ref{sec:web_hier}. The link adding process of FFM for out-going edges is in fact a particular case of the DED sub-score (see Figure \ref{fig:deduction_induction}). This alone indicates that INF\_LOG\_2D and FFM are based on close hierarchical principles. There are nevertheless huge differences between them. FFM does not use a vertex in-edges to determine its out-edges, like INF\_LOG\_2D does through the IND sub-score. INF\_LOG\_2D calculates and rates all edges based on a similarity score, while the FFM randomly accepts and rejects edges as it does not seek faithfulness at a vertex level; it seeks topological coherency at a graph level. Finally, the FFM explores edges far away from the ambassador vertex though various iterations, thanks to its computational simplicity. INF\_LOG\_2D performs only a one step exploration (\ie its a local score), although building a quasi-local version of INF\_LOG\_2D is one of our main lines of future work (see \S\ref{sec:future}). To sum up, while the FFM and INF\_LOG\_2D share a set of precepts, each model uses those for a different purpose: the FFM uses them to define a large, coherent topology model at graph scale, while INF\_LOG\_2D uses them to define a high confidence and exhaustive edge likelihood score applicable at vertex level. In that regard, the good results achieved by both methods on their respective fields partly support the assumptions of the other.


In this paper we also reached an interesting conclusion for the LP field in general. In our analysis of results we defined a simple categorization of SBA based on how these considered evidence: proportionally or accumulatively. Our proposed score INF\_LOG\_2D is actually a hybrid, combining both approaches. While so far results indicate that accumulative solutions are more competitive than proportional ones, the good results of INF\_LOG\_2D open the door to the consideration of hybrid solutions. This is of relevance for the LP field, as it motivates the integration of proportional features into current accumulative scores, with great potential benefits.

From our code implementation and parallelization we derive conclusions for the graph mining and HPC communities. As discussed in \S\ref{sec:experiments}, SBA can be defined as parallel processes without dependencies. Thus, any amount of resources made available to SBA will be used efficiently. Significantly, this feature is consistent in a shared memory or distributed memory settings through the use of OpenMP and OmpSs. This opens the door to the computation of LP to graphs of arbitrary size.

\section{Future Work}\label{sec:future}

From the LP perspective, the goal is to develop more precise prediction scores. Hybrid SBA are a promising family of LP methods that needs to be thoroughly studied. Also, the good performance of INF\_LOG\_2D motivates the design of quasi-local scores based on it, which may achieve even more precise predictions. INF\_LOG\_2D assumes a hierarchical structure, and thus works well on domains which satisfy this model to some degree. Other scores which assume (but do not compute) different underlying models (\eg communities) should also be explored.


Our final line of future work regards large-scale graphs. Our current implementation allows us to compute arbitrarily large graphs with OmpSs. Thanks to that we intend to develop an Internet-wide hyperlink recommender. For that purpose there are questions arising from a distributed memory context that must be considered, such as: How to split the graph data among different physical locations? Which data is to be allocated on each location? When must data be transfered? How is the code parallelized given these new restrictions? By solving this issues we intend to conclude the argument that we are starting with this work: that large-scale link prediction is not a field with a bright future, but instead one with a challenging present.

\section*{Acknowledgements}
This work is partially supported by the Joint Study Agreement no. W156463 under the IBM/BSC Deep Learning Center agreement, by the Spanish Government through Programa Severo Ochoa (SEV-2015-0493), by the Spanish Ministry of Science and Technology through TIN2015-65316-P project and by the Generalitat de Catalunya (contracts 2014-SGR-1051).

\printbibliography

\end{document}